\documentclass[english,aps,prx,reprint,superscriptaddress]{revtex4-1}
\usepackage[T1]{fontenc}
\usepackage[latin9]{inputenc}
\usepackage{geometry}
\usepackage{color}
\usepackage{babel}
\usepackage{verbatim}
\usepackage{textcomp}
\usepackage{amsmath}
\usepackage{amssymb}
\usepackage{graphicx}
\usepackage[unicode=true,pdfusetitle,
 bookmarks=true,bookmarksnumbered=false,bookmarksopen=false,
 breaklinks=false,pdfborder={0 0 0},pdfborderstyle={},backref=false,colorlinks=true]
 {hyperref}
\hypersetup{
 linkcolor=blue, urlcolor=blue, citecolor=blue}

\makeatletter

\newcommand*\LyXThinSpace{\,\hspace{0pt}}

\usepackage{braket}
\usepackage{dsfont}
\usepackage{bm}
\date{}

\makeatother

\begin{document}
\title{Suppressing coherence effects in quantum-measurement based engines}
\author{Zhiyuan Lin}
\thanks{These authors contributed equally to this work.}
\author{Shanhe Su}
\thanks{These authors contributed equally to this work.}
\author{\textcolor{black}{$^{,}$}\textcolor{blue}{$^{\dagger}$}Jingyi Chen}
\author{Jincan Chen}
\email{sushanhe@xmu.edu.cn; jcchen@xmu.edu.cn}

\affiliation{Department of Physics, Xiamen University, Xiamen 361005, People's
Republic of China}
\author{Jonas F. G. Santos}
\affiliation{Centro de Ciências Naturais e Humanas, Universidade Federal do ABC,
Avenida dos Estados 5001, 09210-580 Santo André, São Paulo, Brazil}
\begin{abstract}
The recent advances in the study of thermodynamics of microscopic
processes have driven the search for new developments in energy converters
utilizing quantum effects. We here propose a universal framework to
describe the thermodynamics of a quantum engine fueled by quantum
projective measurements. Standard quantum thermal machines operating
in a finite-time regime with a driven Hamiltonian that does not commute
in different times have the performance decreased by the presence
of coherence, which is associated with a larger entropy production
and irreversibility degree. However, we show that replacing the standard
hot thermal reservoir by a projective measurement operation with general
basis in the Bloch sphere and controlling the basis angles suitably
could improve the performance of the quantum engine \textcolor{black}{as
well as decrease the entropy change during the measurement process.}\textcolor{red}{{}
}Our results go in direction of a generalization of quantum thermal
machine models where the fuel comes from general sources beyond the
standard thermal reservoir.
\end{abstract}
\maketitle

The current development of new devices based on quantum effects has
shown that a well-formulated understanding of energy conversion is
required. This has been clearly demonstrated through the considerable
number of theoretical and experimental advances in quantum thermodynamics
\cite{Ro=0000DFnagel2016,Kosloff2013,chen2016,chen2002}. It must
be highlighted recent progresses in quantum fluctuation relations
\cite{Batalh=0000E3o2014,=0000C5berg2018,Micadei2020}, thermodynamics
uncertainty relations \cite{Timpanaro2019,Lee2021,Sacchi2021}, as
well as the study of quantum protocols involving non-equilibrium thermal
baths \cite{Zhang2020,Huang2012,Scully2003,Rodrigues2019}, strong
coupling regime \cite{Cresser2021,Strasberg2016}, and non-Markovian
effects \cite{Breuer2009,Laine2010}.

Encompassing the above mentioned results, the study of quantum thermal
machines (QTMs) is useful to investigate how specific driving protocols
or non-classical baths impact the performance of thermodynamic cycles,
i.e, the work extraction (cooling rate) for engine (refrigerator)
configurations. In this direction, it is well known that coherence
in the energy basis of the working substance leads to the increase
of entropy production along the cycle, resulting in the degradation
of the performance \cite{Camati2019,Santos2019,Francica2019}. The
theoretical design of quantum machines employing other models of bath
beyond the standard Markovian thermal reservoir is also another interesting
aspect that has been addressed. For instance, it could be highlighted
the use of non-Markovian thermal bath \cite{Abiuso2019,Shirai2021,Guarnieri2016,Camati2020},
finite-size environment \cite{Pozas-Kerstjens2018}, and squeezed
thermal baths \cite{Assis2020,Klaers2017,Ro=0000DFnagel2014}. For
a quantum thermal machine operating in a finite-time Otto cycle, reaching
the limit efficiency with a non-zero extracted power is motivated
by realistic applications. From an experimental point of view, realizations
of a single-spin quantum engine have been performed in nuclear magnetic
resonance (NMR) \cite{Peterson2019} and in an ensemble of nitrogen
vacancy centers in diamond \cite{Klatzow2019}, both pointing out
the generation of coherence in the energy basis of the working substance.
Also, an experimental verification of the fluctuation relation for
work and heat in a quantum engine has been recently reported in Ref.
\cite{Denzler2021}, showing how correlations between work and heat
affect the performance of a finite-time quantum Otto engine.

\textcolor{black}{Apart the standard QTM where the fuel is supplied
by a thermal reservoir, a mo}re general type of QTM was proposed by
Szilard \cite{Szilard1929} and differs from the former models in
the sense that energy is extracted from a single heat bath by using
a feedback mechanism well-known as Maxwell demon \cite{Leff1990,Elouard2017},
putting in evidence that information can effectively work as a fuel
in general QTM models \cite{Park2013,Toyabe2010}. In order to have
information flowing into the system in a QTM, we necessarily need
to measure the system employing a specific protocol. Since a measurement
performed on a quantum system generally alters its state, recent advances
have proposed a new QTM type, well-known as meas\textcolor{black}{urement-driven
engines, in which the measurement protocol itself acts as a fuel and
allows for work extraction \cite{Yi2017,Ding2018,Behzadi2020,Elouard2018,Bresque2021,Brandner2015}.
For standard projective measurements, the system state is entirely
collapsed to a specific eigenstate. There is also a special activity
concerning the so-called weak measurement \cite{Aharonov1988,Duck1989,Flack2014},
whose system state is only partially perturbed \cite{Hosten2008,Lundeen2011,Monroe2021,Jacobs2014,Wiseman2010}.
A natural question that arises in thermodynamic cycles is how to unveil
the role played by quantum measu}rements in the energy conversion
and the possibility to suitably engender quantum measurements to enhance
the performance of QTMs. In particular for a quantum Otto-like cycle,
the finite-time driven unitary strokes in general induce transition
between the energy eigenstates of the working substance \cite{Camati2019,Zagoskin2012},
resulting in an efficiency bellow the Otto limit. However, replacing
the standard hot reservoir by a quantum measurement protocol provides
another source for transitions which depend essentially on the measurement
basis. Thus, from a point of view of the friction induced by transitions
between the eigenstates of the working substance \cite{Rezek2010,Feldmann2006,Kosloff2002},
quantum projective measurements with suitable basis on the Bloch sphere
could be understood as a kind of quantum lubricant if it attenuates
the degradation effect due to the presence of coherence.

In this work, we are interested in addressing how quantum fluctuations
associated with finite-time driven unitary processes and quantum measurements
affect the performance of a quantum engine. For this purpose, we consider
a modified version of the quantum Otto cycle where the standard hot
thermal reservoir is replaced by a quantum projective measurement,
providing a different mechanism to fuel the cycle. In order to understand
the role played by two kind of quantum fluctuations, we present explicitly
expressions for the thermodynamic quantities characterizing the engine,
i.e, the extracted work, heat absorbed by the working substance due
to the measurement protocol, heat released to the cold source, and
efficiency as functions of transition probabilities related to the
finite-time driven Hamiltonian and the measurement scheme. Considering
a numerical simulation with parameters employed in NMR setup \cite{Oliveira_Book2007},
we show that having a sufficient control over the choice of the measurement
basis, it is possible to enhance the work extraction and then the
performance of the quantum en\textcolor{black}{gine. Since the irreversibility
of the cycle is associated with the increase of entropy, we also show
that the maximum values of efficiency and extracted work are reached
very close to the minimum value of entropy change during the measurement
protocol.}

\textit{The cycle of }\textit{\textcolor{black}{the engine based on
quantum measurement}}\textit{.---}The quantum measurement engine
employs a particle of spin 1/2 (single qubit) as the working substance.
To complete one operating cycle, the engine goes through four strokes,
including two adiabatic (unitary) processes, a quantum measurement
process, and a thermalization process, as sketched in Fig. 1. The
working substance is initially prepared in thermal equilibrium with
a heat bath at positive inverse temperature $\beta$, such that its
state at time $t=0$ is given by $\rho_{1}=e^{-\beta H_{1}}/Z_{1}$\cite{Quan2007,Chen2018},
where $H_{1}=\frac{\hbar\omega}{2}\sigma_{z}$ is the initial Hamiltonian,
$Z_{1}=\mathrm{Tr}\left(e^{-\beta H_{1}}\right)$ is the partition
function, $\hbar$ is the reduced Planck constant, $\omega$ denotes
the resonance frequency, and $\sigma_{i}\left(i=x,y,z\right)$ are
the Pauli matrices.

Durning the adiabatic stroke at stage I, a time-modulated radiofrequency
field generates a time-dependent Hamiltonian $H_{I}(t)=\frac{\hbar\omega}{2}\left(\cos\frac{\pi t}{2\tau}\sigma_{z}+\textrm{sin}\frac{\pi t}{2\tau}\sigma_{x}\right)$,
which clearly does not commute in different times, implying the generation
of coherence in the energy basis of the working substance \cite{Camati2019}.
At $t=\tau$, the working substance changes into state $\rho_{2}=U_{\tau,0}\rho_{1}U_{\tau,0}^{\dagger}$
with $U_{\tau,0}=\mathcal{T}e^{-\frac{i}{\hbar}\int_{0}^{\tau}H_{I}\left(t\right)dt}$
being the the time-evolution operator and $\mathcal{T}$ being the
time-ordering operator. The work performed on the spin $\left\langle W_{1}\right\rangle =\textrm{Tr}\left(\rho_{2}H_{2}-\rho_{1}H_{1}\right)$,
where $H_{2}=\frac{\hbar\omega}{2}\sigma_{x}$ represents the final
Hamiltonian. After the quantum measurement at stage II, the state
of the spin is updated to $\rho_{3}=\sum_{k}\pi_{k}\rho_{2}\pi_{k}$
\cite{Manzano2018,Lloyd1997}, where $\pi_{k}=\left|\chi_{k}\right\rangle \left\langle \chi_{k}\right|$
denotes the orthogonal projector associated with the measurement basis
$\left|\chi_{1}\right\rangle =e^{-i\varphi}\sin\frac{\alpha}{2}\left|\uparrow\right\rangle -\cos\frac{\alpha}{2}\left|\downarrow\right\rangle $
and $\left|\chi_{2}\right\rangle =\cos\frac{\alpha}{2}\left|\uparrow\right\rangle +e^{i\varphi}\sin\frac{\alpha}{2}\left|\downarrow\right\rangle $.
In the Bloch sphere representation, $\alpha$ and $\varphi$ are,
respectively, the colatitude with respect to the z-axis and the longitude
with respect to the x-axis. The total fuel energy provided by quantum
measument $\left\langle Q_{M}\right\rangle =\textrm{Tr}\left[H_{2}\left(\rho_{3}-\rho_{2}\right)\right]$.
Durning the adiabatic process at stage III, the driving Hamiltonian
$H_{II}(t)=H_{I}(2\tau-t)$ with $t\in\left[\tau,2\tau\right]$. The
final state of the second adiabatic stroke $\rho_{4}=V_{2\tau,\tau}\rho_{3}V_{2\tau,\tau}^{\dagger}$
, where the unitary operator $V_{2\tau,\tau}=\mathcal{T}e^{-\frac{i}{\hbar}\int_{\tau}^{2\tau}H_{II}\left(t\right)dt}$
. The work performed by the exteral field $\left\langle W_{2}\right\rangle =\textrm{Tr}\left(\rho_{4}H_{1}-\rho_{3}H_{2}\right)$.
The spin returns to the initial Gibbs state $\rho_{1}$ for a thermalization
process at stage IV. The heat $\left\langle Q_{T}\right\rangle $
flowing into the spin from the bath $\left\langle Q_{T}\right\rangle =\textrm{Tr}\left(H_{1}\rho_{1}-H_{1}\rho_{4}\right)$. 

A more detailed description of the cycle of the quantum measurement
engine is shown in Supplementary I. One is capable of proving that
$\left\langle Q_{T}\right\rangle $ is always negative, meaning that
energy is actually flowing from the working substance into the heat
bath (Supplementary II). The thermodynamic cycle is impossible to
convert the heat from a single source into work without any other
effect, satifying Kelvin's statement of the second law of thermodynamics. 

\noindent 
\begin{figure}
\noindent \begin{centering}
\includegraphics[scale=0.25]{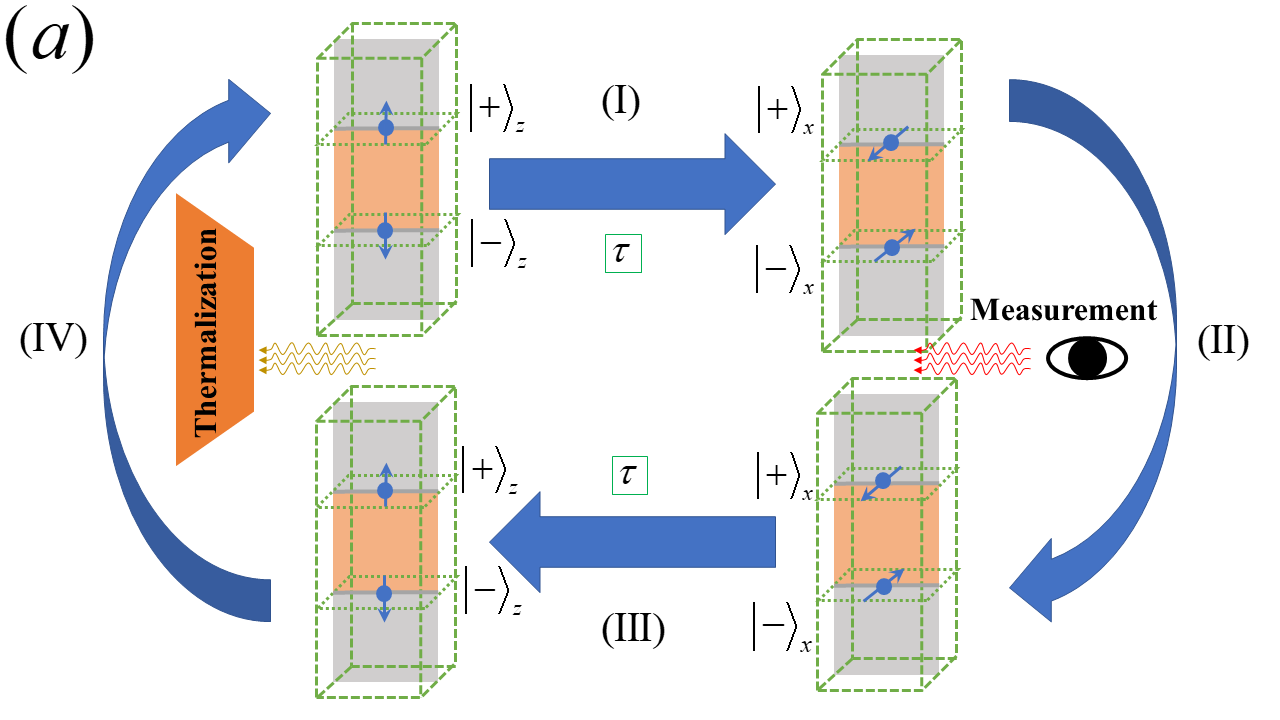}
\par\end{centering}
\noindent \begin{centering}
\includegraphics[scale=0.25]{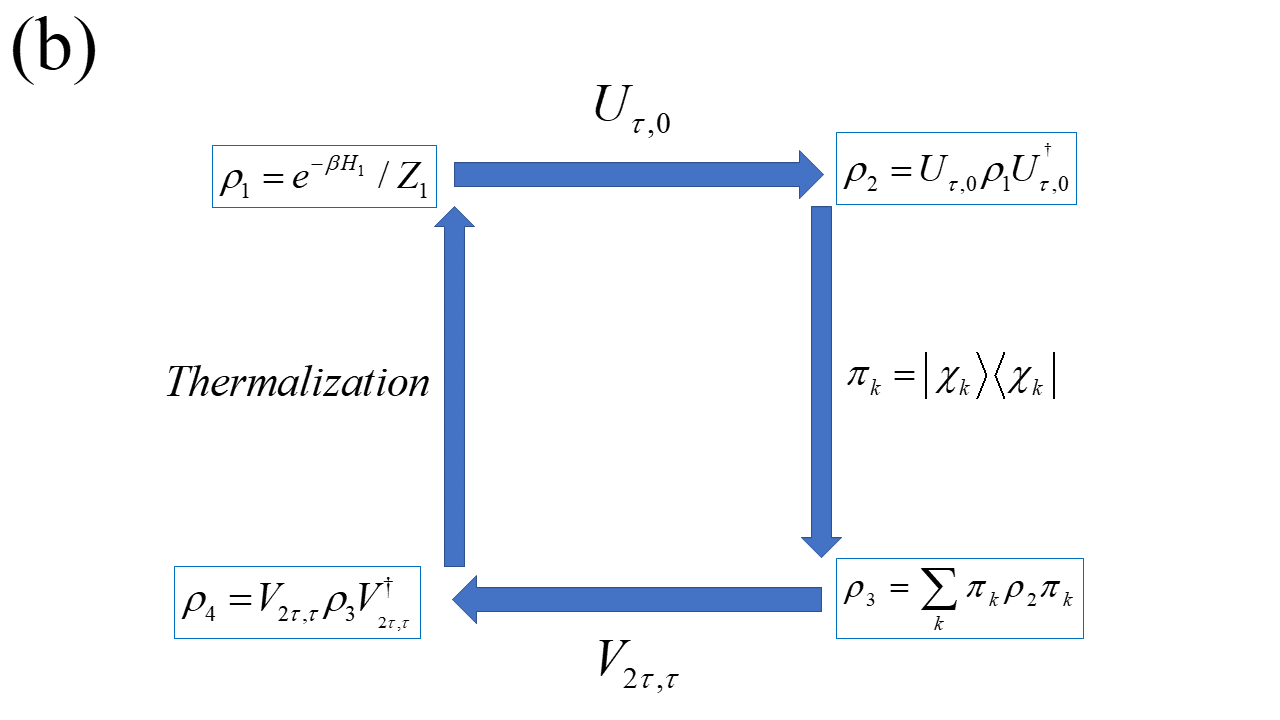}
\par\end{centering}
\caption{\textcolor{black}{Illustration of the engine based on quantum measurement.
(a) The working substance starts in thermal equilibrium with the heat
bath. The first stroke is an adiabatic process mediated by a time-dependent
Hamiltonian. In the second stroke, an instantaneous projective measurement
is performed on the working substance, projecting the single-qubit
onto the basis $\left\{ \left|\chi_{1}\right\rangle ,\left|\chi_{2}\right\rangle \right\} $.
The third stroke is again a unitary evolution using a time-dependent
Hamiltonian. In the fourth stroke, the working substance relaxes to
the initial thermal equilibrium state. (b) The evolution of the density
matrix during the cycle.}}
\end{figure}

\textit{The roles of transition probabilities in the performance of
the engine.---}The net work done by the external agent is 

\begin{equation}
\langle W\rangle=\langle W_{1}\rangle+\langle W_{2}\rangle=\hbar\omega\left[\xi-\left(\delta-\gamma\right)\left(1-2\zeta\right)\right]\tanh\left(\frac{\beta\hbar\omega}{2}\right),\label{eq:work}
\end{equation}
where $\zeta=\left|\left\langle \chi_{2}\right|U_{\tau,0}\left|-\right\rangle _{z}\right|^{2}\left(\delta=\left|\left\langle \chi_{2}\right|U_{\tau,0}|-\rangle_{x}\right|^{2}\right)$
is the transition probability between the basis state $\left|\chi_{2}\right\rangle $
of measurement and the ground eigenstate $\left|-\right\rangle _{z}\left(\left|-\right\rangle _{x}\right)$
of $H_{1}\left(H_{2}\right)$, $\xi=\left|_{x}\left\langle +\right|U_{\tau,0}\left|-\right\rangle _{z}\right|^{2}$
is the transition probability between the excited eigenstate $\left|+\right\rangle _{x}$
of $H_{2}$ and the ground eigenstate $\left|-\right\rangle _{z}$
of $H_{1}$ due to the unitary evolution at stage I, and $\gamma=\left|_{z}\left\langle +\right|V_{2\tau,\tau}\left|\chi_{1}\right\rangle \right|^{2}$
is the transition probability between the excited eigenstate $\left|+\right\rangle _{z}$
of $H_{1}$ and the basis state $\left|\chi_{1}\right\rangle $ of
measurement due to the unitary evolution at stage III. The transition
probabilities embody the influence of quantum fluctuations and satisfy
the principle of microreversibility (Supplementary IV) \cite{Campisi2011,Ehrich2020}.
For the purpose of extracting work from the engine, we must have $\langle W\rangle<0$.
The fuel energy provided by the measurement process reads

\begin{equation}
\langle Q_{M}\rangle=\frac{\hbar\omega}{2}\left[\left(1-2\xi\right)-\left(1-2\delta\right)\left(1-2\zeta\right)\right]\tanh\left(\frac{\beta\hbar\omega}{2}\right).\label{eq:measurement heat}
\end{equation}

The heat released by the working substance to the cold thermal reservoir
in the thermalization process (stage IV) is written as

\begin{equation}
\langle Q_{T}\rangle=\frac{\hbar\omega}{2}\left[\left(1-2\gamma\right)\left(1-2\zeta\right)-1\right]\tanh\left(\frac{\beta\hbar\omega}{2}\right).\label{cold heat}
\end{equation}
Note that since $0\leq\gamma\leq1$ and $0\leq\zeta\leq1$, the inequality
$\frac{1}{\zeta}+\frac{1}{\gamma}\geq2$ ensures that $\langle Q_{T}\rangle\leq0$
(Supplementary II). 

The performance of the quantum engine is dictated by the efficiency
defined as $\eta=-\langle W\rangle/\langle Q_{M}\rangle=1+\langle Q_{T}\rangle/\langle Q_{M}\rangle$,
where the second expression comes from the first law of thermodynamics.
Using Eqs. (\ref{eq:measurement heat}) and (\ref{cold heat}) the
efficiency reads

\begin{equation}
\eta=1-\frac{\left(1-2\gamma\right)\left(1-2\zeta\right)-1}{\left(1-2\delta\right)\left(1-2\zeta\right)-\left(1-2\xi\right)}.\label{eq:eff}
\end{equation}
The efficiency is limited by $0\leq\eta\leq1$ because of the constraints
$\langle Q_{T}\rangle\leq0\leq\langle Q_{M}\rangle$ and $\left|\langle Q_{T}\rangle\right|\leq\left|\langle Q_{M}\rangle\right|$.\textcolor{red}{{}
}\textcolor{black}{With the set of Eqs. (\ref{eq:work}-\ref{eq:eff}),
all the quantum fluctuations induced by the time-dependent Hamiltonian
and the measurement protocols are being taken into account, and affect
the performance of the engine. In particular, we will show how quantum
fluctuations arising from the measurement protocols could be employed
(the choice of the measurement angles) in order to suppress the degradation
effect due to the coherence, then increasing the efficiency of the
cycle.}

\textit{\textcolor{black}{Numerical simulation.---}}\textcolor{black}{In
this section, we illustrate our results with a numerical simulation
with feasible parameters employed in NMR setup. Firstly, we note that
when $\alpha=\pi/2$ and $\varphi=0$, the measurement basis commutes
with the eigenstate basis of $H_{2}$. Therefore, no energy is delivered
to the working substance, resulting in zero efficiency and null extracted
work. }The projective quantum measurement inevitably alters the mean
energy of the observed system if the measurement basis does not commute
with the bare energy basis. This is the reason why quantum measurement
is able to supply a continuous source of energy similar to the combustion
of a fuel. Formally, the measurement basis corresponding to measuring
the single-qubit is a trigonometric function of the colatitude $\alpha$
and longitude $\varphi$ . From Fig. (2), we conclude that $\alpha$
and $\varphi$ are two crucial independent parameters to determine
the performance of the engine. The working areas of the engine are
divided into two separate parts, i.e., $0\leq\varphi\leq\pi$ and
$\pi\leq\varphi\leq2\pi$. For $0\leq\varphi\leq\pi$, maximum work
output $\left(-\langle W\rangle\right)_{\text{max}}$ and efficiency
$\eta_{\text{max}}$ appear at different measurement directions. As
depicted in Fig. (2), $-\langle W\rangle$ qualitatively peaks at
$\alpha_{W}=1.39$ and $\varphi_{W}=2.05$, while $\eta$ reaches
its maximum at $\alpha_{\eta}=1.45$ and $\varphi_{\eta}=2.53$. To
obtain a maximum attainable efficiency at a given extracted work,
the optimal ranges of the extracted work and efficiency must be constrained
by $\alpha_{W}\leq\alpha\leq\alpha_{\eta}$ and $\varphi_{W}\leq\varphi\leq\varphi_{\eta}$.
\textcolor{black}{For} $\pi\leq\varphi\leq2\pi$, the distributions
of \textminus $\left\langle W\right\rangle $ and $\eta$ satisfy
antisymmetry with the axis $\alpha=\pi/2$ and translational invariance,
i.e., $-\left\langle W\right\rangle \left(\alpha,\varphi\right)=-\left\langle W\right\rangle \left(\pi-\alpha,\varphi+\pi\right)$
and $\eta\left(\alpha,\varphi\right)=\eta\left(\pi-\alpha,\varphi+\pi\right)$. 

\noindent 
\begin{figure}
\noindent \begin{centering}
\includegraphics[scale=0.3]{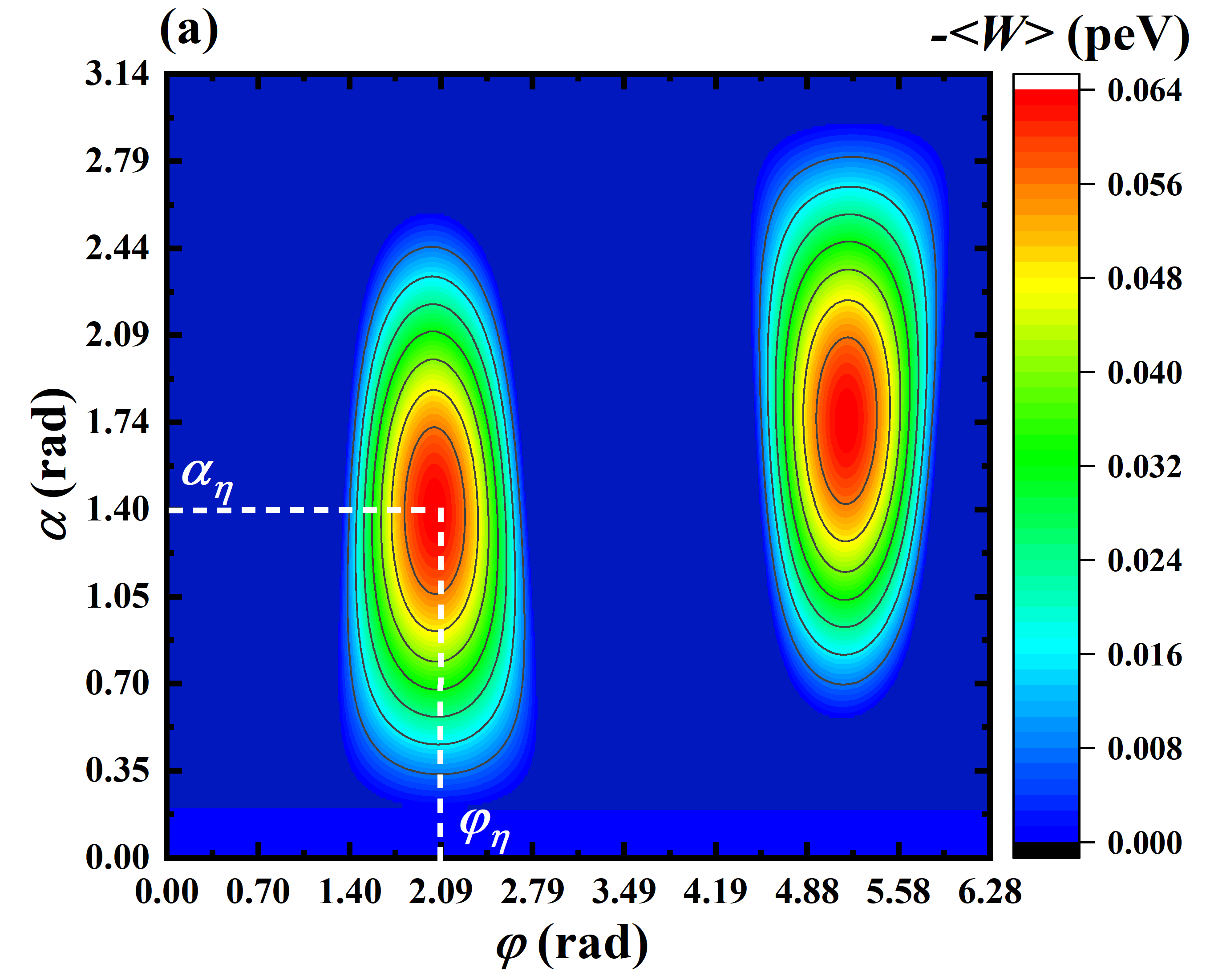}
\par\end{centering}
\noindent \begin{centering}
\includegraphics[scale=0.3]{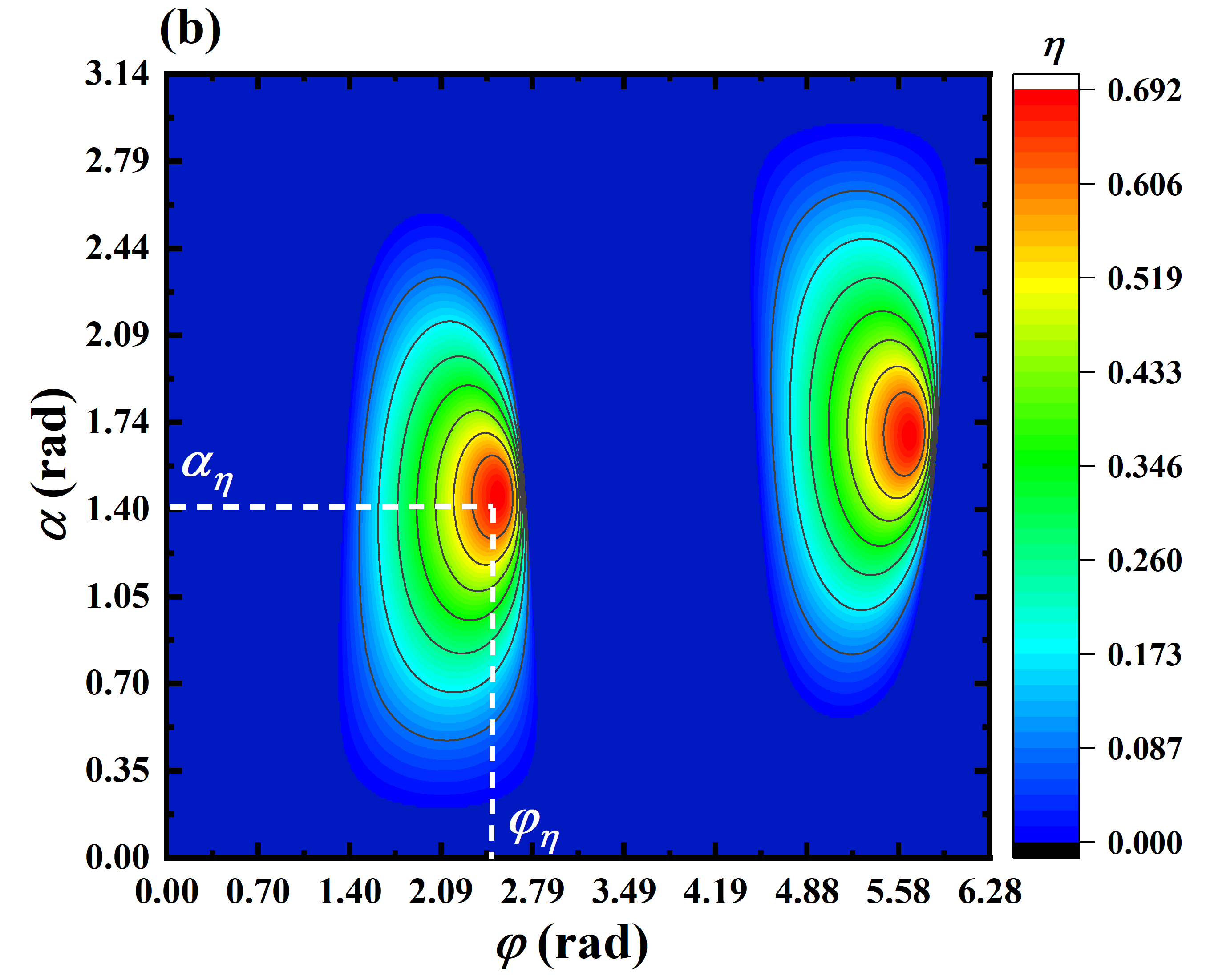}
\par\end{centering}
\includegraphics[scale=0.5]{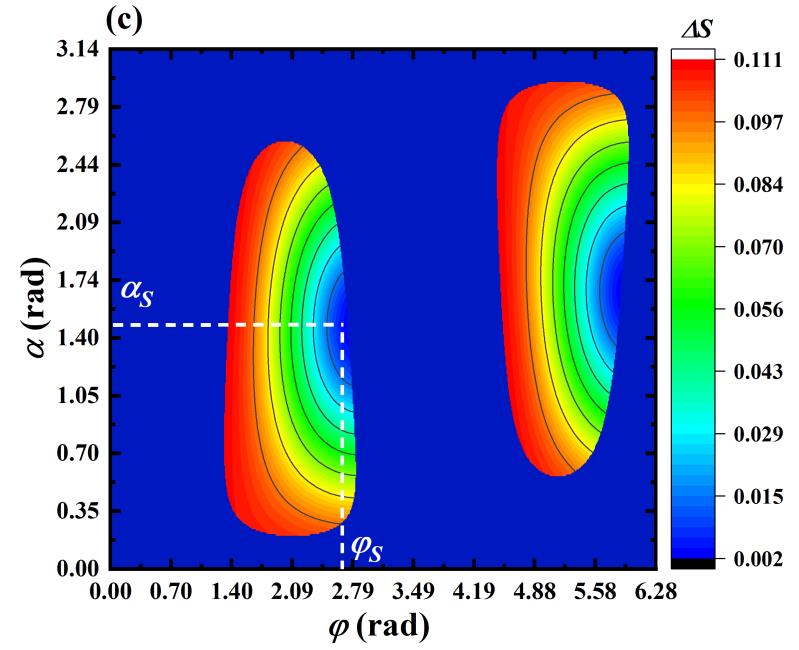}\label{fig2}

\caption{\textcolor{black}{Performance of the engine based on quantum measurement.
(a) The extracted work $-\langle W\rangle$, (b) efficiency $\eta$,
and (c) entropy change $\Delta S$ during the measurement process
(stage II) varying with the colatitude $\alpha$ and longitude $\varphi$
on the Bloch sphere, where $\hbar\omega=1\textrm{peV}$, $\tau=10\textrm{\ensuremath{\mu}s}$,
and $\beta=1/\left(\hbar\omega\right)$. These values are used unless
otherwise mentioned specifically in the following discussion.}}
\end{figure}

To understand the physics behind the enhancement of $-\langle W\rangle$
and $\eta$, we write $-\langle W\rangle$ and the quantum fuel $\langle Q_{M}\rangle$
in terms of the occupation probabilities, i.e.,

\begin{align}
\langle W\rangle & =-\frac{\hbar\omega}{2}\left(\Delta p_{1}-\Delta p_{2}+\Delta p_{3}-\Delta p_{4}\right)\nonumber \\
\langle Q_{M}\rangle & =\frac{\hbar\omega}{2}\left(\Delta p_{2}-\Delta p_{3}\right),\label{performance_probability}
\end{align}
where $\Delta p_{1}=-\textrm{Tr}\left(\rho_{1}\sigma_{z}\right)=\tanh\left(\beta\hbar\omega/2\right)$,
$\Delta p_{2}=-\textrm{Tr}\left(\rho_{2}\sigma_{x}\right)=\Delta p_{1}\left(1-2\xi\right)$,
$\Delta p_{3}=\left\langle \chi_{1}\right|\rho_{3}\left|\chi_{1}\right\rangle -\left\langle \chi_{2}\right|\rho_{3}\left|\chi_{2}\right\rangle =\Delta p_{1}\left(1-2\delta\right)\left(1-2\zeta\right)$,
and $\Delta p_{4}=-\textrm{Tr}\left(\rho_{4}\sigma_{z}\right)=\Delta p_{1}\left(1-2\gamma\right)\left(1-2\zeta\right)$
are, respectively, the difference in the occupation probability between
the ground and excited states of each quantum state. As a result,
the efficiency is simplified as

\begin{equation}
\eta=1-\left(\Delta p_{1}-\Delta p_{4}\right)/\left(\Delta p_{2}-\Delta p_{3}\right),
\end{equation}
which is completely determined by the probability changes caused by
the transition coefficients. A conventional quantum engine carries
out an adiabatic compression/expansion through expanding/reducing
the gap between energy levels of the Hamiltonian. The emergence of
measurement indicates that the purpose of extracting work may be achieved
without changing the energy level spacing.

\textcolor{black}{Apart the energetic exchanges in a quantum cycle,
entropic quantities are inherently associated to the irreversibility
of the cycle. In this sense, the entropy change of the working substance
during the measurement process (state II) directly affects the performance
of the engine. Figure 2(c) depicts the entropy change $\Delta S=S(\rho_{3})-S(\rho_{2})$
during the measurement process, where $S(\rho_{i})=-Tr\left(\rho_{i}\ln\rho_{i}\right)$
is the von Neumann entropy of a given state. The entropy change is
minimum at $\left(\alpha_{S},\varphi_{S}\right)=\left(1.46,2.74\right)$,
given the parameters considered in the numerical simulation. The
region of large efficiencies matches very well with the region of
small entropy changes during the measurement process. This is in agreement
with the prediction that decreasing the irreversibility increases
the performance of quantum engines. In this sense, we could argue
that by choosing an suitable basis for the measurement protocol, it
works effectively as a kind of quantum friction, suppressing the degradation
effect due to the coherence generated in the stages I and III. }The
equalities $S(\rho_{1})=S(\rho_{2})$ and $S(\rho_{3})=S(\rho_{4})$
hold because of the invariant of von Neumann entropy under a unitary
evolution. In the fourth stroke of the cycle, one can then confirm
that the entropy change $S(\rho_{1})-S(\rho_{4})$ caused by the thermalization
process is equal to $-\Delta S$. 

\begin{figure}
\noindent \begin{centering}
\includegraphics[scale=0.36]{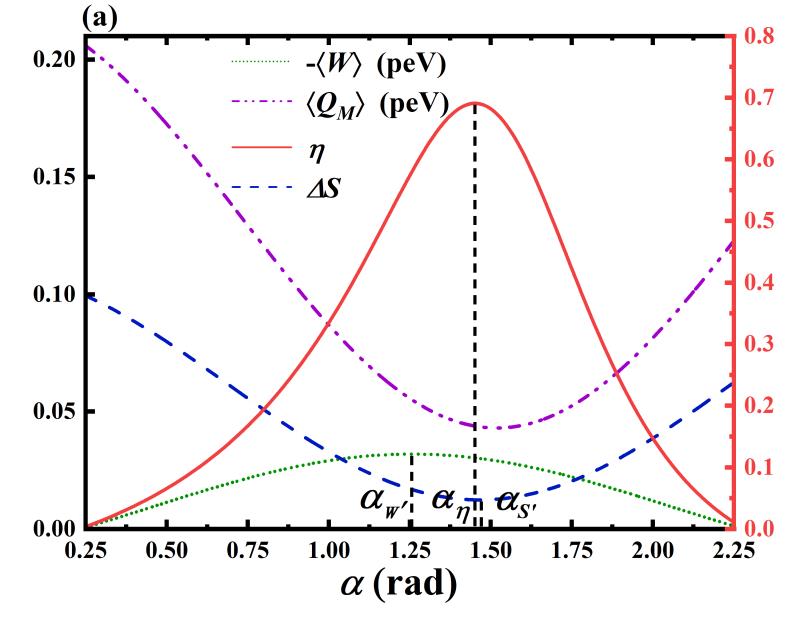}\includegraphics[scale=0.2]{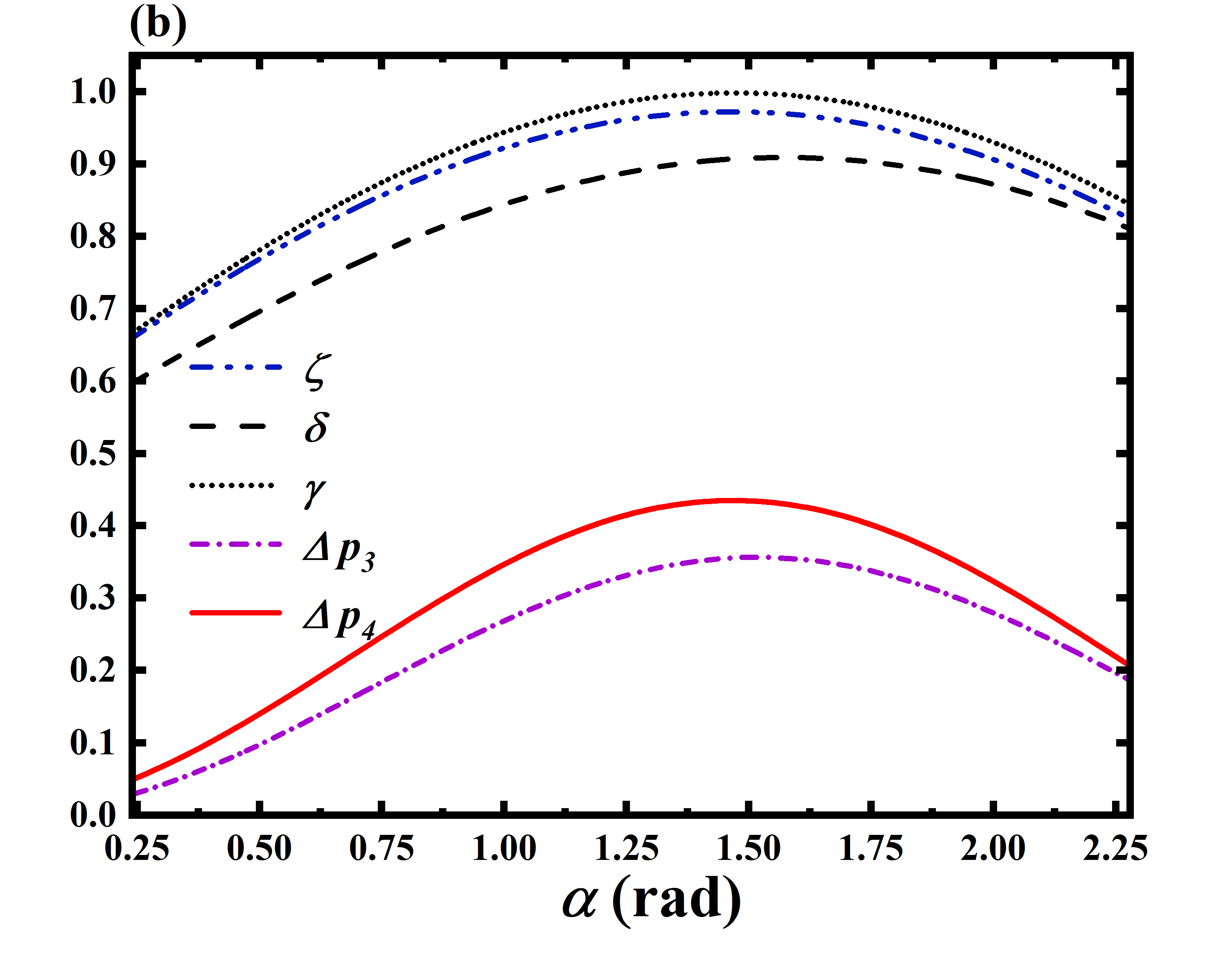}
\par\end{centering}
\noindent \begin{centering}
\includegraphics[scale=0.36]{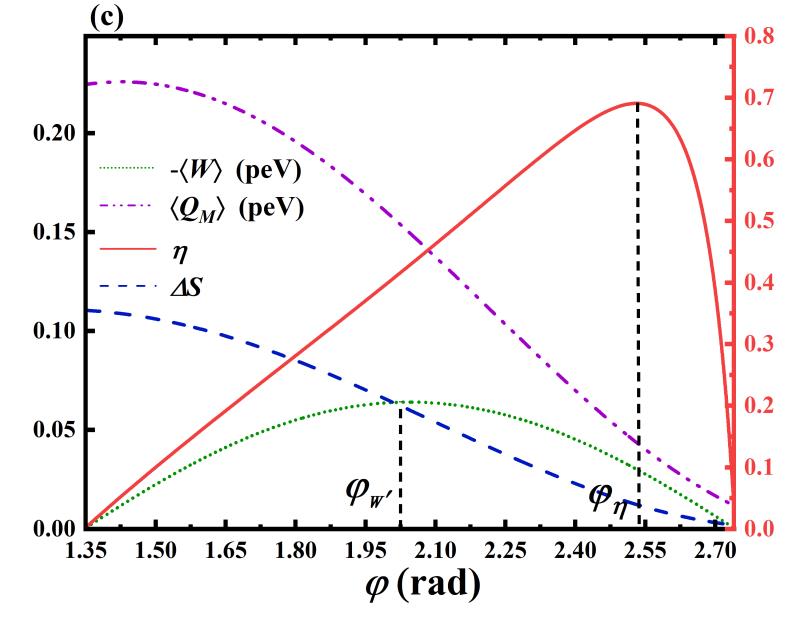}\includegraphics[scale=0.2]{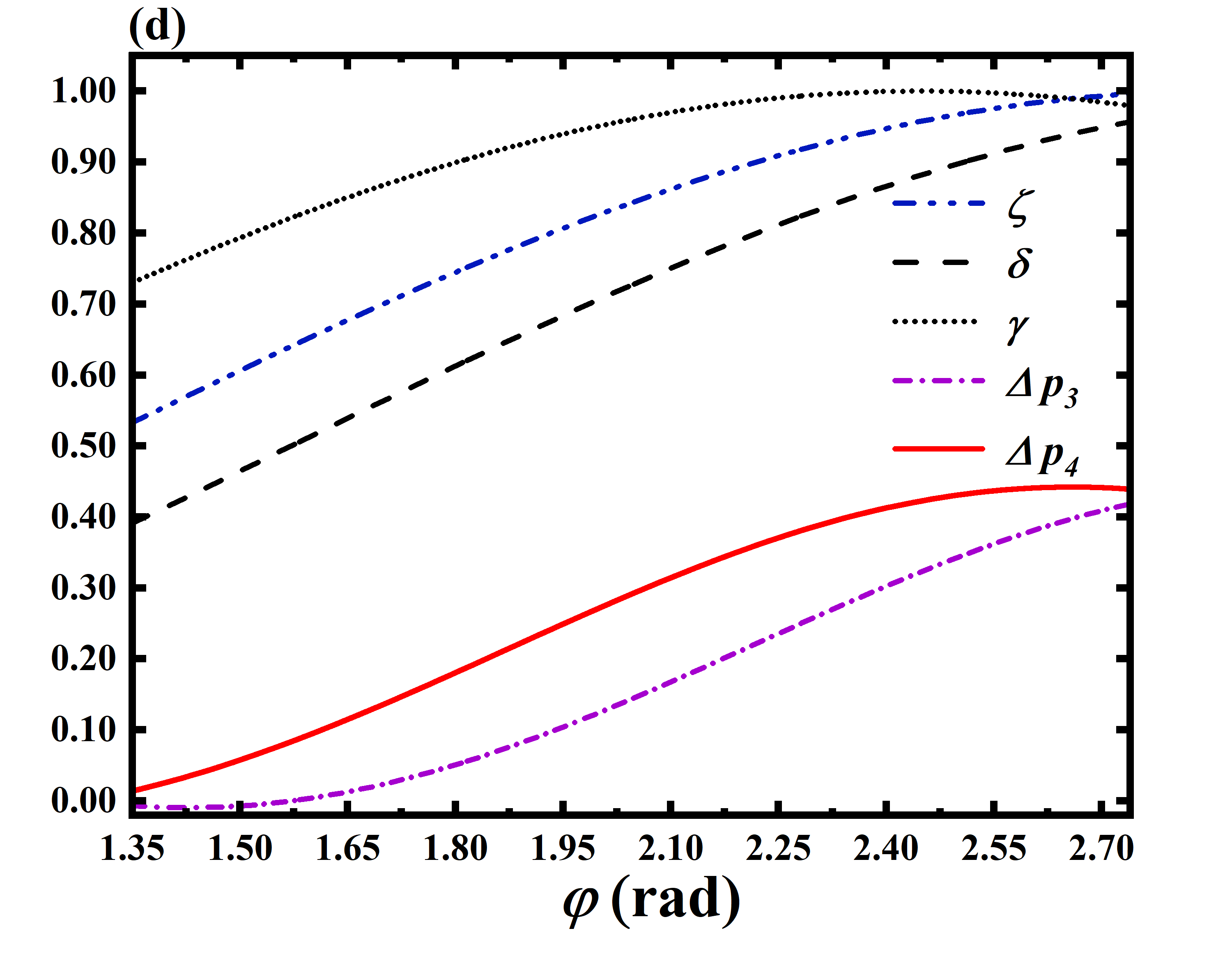}
\par\end{centering}
\caption{For a given value $\varphi_{\eta}$ of $\varphi$, the curves of (a)
the work output $-\langle W\rangle$ , input energy $\langle Q_{M}\rangle$,
efficiency $\eta$, and the entropy change during the measurement
process $\Delta S$ and (b) the transition probabilities $\zeta$,
$\delta$, and $\gamma$ and the differences $\varDelta p_{3}$ and
$\varDelta p_{4}$ in probability distributions between the ground
and excited states varying with the colatitude $\alpha$. For a given
value $\alpha_{\eta}$ of $\alpha$, the curves of (c) the work output
$-\langle W\rangle$ , input energy $\langle Q_{M}\rangle$, efficiency
$\eta$, and the entropy change during the measurement process $\Delta S$
and (d) the transition probabilities $\zeta$, $\delta$, and $\gamma$
and the difference $\varDelta p_{3}$ and $\varDelta p_{4}$ in probability
distributions between the ground and excited states varying with the
longitude $\varphi$ . In Figs. (a) and (c), the left vertical axis
shows values for $-\langle W\rangle$ and $\langle Q_{M}\rangle$,
while the corresponding scale of $\eta$ is on the right vertical
axis.}
\end{figure}

In Fig. 3 (a), $-\langle W\rangle$ $\left(\eta\right)$ reaches its
limit at $\alpha_{W^{\prime}}=1.25$ $\left(\alpha_{\eta}=1.45\right)$.
However, the input energy $\langle Q_{M}\rangle$ is relatively small
at the points of $\alpha_{W^{\prime}}$ and $\alpha_{\eta}$. \textcolor{black}{We
also see the behavior of the entropy change during the measurement
process, evidencing that it is considerably small for high values
of the efficiency. }The measurement based engine could generate a
greater amount of work at low cost by optimizing the angles of the
measurement basis. As the colatitude $\alpha$ changes, the probability
changes $\varDelta p_{3}$ and $\varDelta p_{4}$ are the only factors
that alter the useful extracted energy and the total energy input.
To go a step further, the variations of $\varDelta p_{3}$ and $\varDelta p_{4}$
lie on the term associated with $\zeta$, $\delta$, and $\gamma$,
as shown in Fig. 3(b). Note that the parameters $\xi$, $\Delta p_{1}$,
and $\Delta p_{2}$ remain constant at any given driving time and
are not showing up in the graph. The transition probabilities $\zeta$,
$\delta$, and $\gamma$ depending on states $\chi_{1}$ and $\chi_{2}$
contain all information about how quantum measurement plays an important
role in thermodynamics. Figure 3(b) reveals that $1-2\zeta$, $1-2\delta$,
and $1-2\gamma$ are convex functions of $\alpha$, because their
derivatives are monotonically non-decreasing. Numerical results also
show that the product of $1-2\delta\left(1-2\gamma\right)$ and $1-2\zeta$
is a concave function, resulting in the existence of a local maximum
of $\Delta p_{3}\left(\Delta p_{4}\right)$. When $\Delta p_{3}$
takes the peak value, the working substance has the largest probability
of being found in the ground state after the measurement and the input
energy $\langle Q_{M}\rangle$ would be minimum. It is also observed
that the gap between $\Delta p_{3}$ and $\Delta p_{4}$ at $\alpha_{W^{\prime}}$
determines the upper bound of $-\langle W\rangle$.

Finally, we examine how the longitude $\varphi$ of the measurement
basis influences the performance. Figure 3(c) shows that $-\langle W\rangle$
$\left(\eta\right)$ qualitatively peaks at $\varphi_{W^{\prime}}=2.05$
$\left(\varphi_{\eta}=2.53\right)$. In the small-$\varphi$ regime
$\left(\varphi<\varphi_{W^{\prime}}\right)$, the difference between
$\Delta p_{4}$ and $\Delta p_{3}$ is enhanced as the increase of
$\varphi$ is attempting to raise $\left|1-2\zeta\right|$ {[}Fig.
3(d){]}. However, in the large-$\varphi$ regime $\left(\varphi\text{>}\varphi_{W^{\prime}}\right)$,
the decrease of the discrepancy $\gamma-\delta$ determines the reduction
of $\Delta p_{4}-\Delta p_{3}$. $-\langle W\rangle$ and $\Delta p_{4}-\Delta p_{3}$
have a strongly positive, linear relationship. Overall, $\langle Q_{M}\rangle$
rapidly decreases with the growth of $\varphi$, since the increase
of the transition probabilities $\delta$ and $\zeta$ with respect
to increasing $\varphi$ whittles down $\Delta p_{3}$ . The above
analysis reveals that the engine under the finite-time adiabatic driving
regime realizes the work extraction without changing the spin transition
frequency. The angles of measurement basis on the Bloch sphere determine
the upper limits on the average work output and efficiency.\textcolor{red}{{}
}\textcolor{black}{Again, Fig. 3(c) also shows that the maximum value
for the efficiency is found for small values of entropy change during
the measurement process.}

\textit{Conclusions.---}Recent advances have shown the possibility
of harnessing the energy provided by quantum measurements. We here
develop a measurement based single-qubit quantum engine and show how
quantum measurement is able to work as a fuel in a quantum cycle.
The present model employs a Hamiltonian that does not commute in different
times, thus generating coherence in the energy basis of the working
substance and then decreasing the performance of the quantum engine.
By assuming a sufficient control under the measurement basis angles
$\alpha$ and $\varphi$, we are able to circumvent the degradation
effect due to coherence and increase the extracted work and efficiency.\textcolor{black}{{}
This is verified by comparing the entropy change during the measurement
process (stage II) and the efficiency of the engine, showing that
high values of efficiency coincide with small values of entropy change.
Thus, we could argue that a suitable choice of the measurement basis
effectively works as a kind of quantum lubrication, since it suppresses
the effect of the coherence produced in stage I and III.}

Our results indicate that quantum measurement can be useful to build
quantum thermodynamic cycles beyond the standard ones with two thermal
baths. Besides, the numerical simulation considers parameters usually
employed in NMR, which opens the possibility to experimentally test
the measurement based single-qubit quantum engine. We hope that this
work can help to unveil the role played by measurement in quantum
thermodynamics and its applications.

\textit{Acknowledgements.---}This work has been supported by the
National Natural Science Foundation of China (Grant No. 11805159 and
12075197) and the Natural Science Foundation of Fujian Province (No.
2019J05003). Jonas F. G. Santos acknowledges São Paulo Research Grant
No. 2019/04184-5.


\begin{thebibliography}{10}
\bibitem{Ro=0000DFnagel2016}J. Roßnagel, S. T. Dawkins, K. N. Tolazzi,
O. Abah, E. Lutz, F. Schmidt-Kaler, and K. Singer, A single-atom heat
engine, Science \textbf{352}, 325 (2016).

\bibitem{Kosloff2013}R. Kosloff, Quantum thermodynamics: A dynamical
viewpoint, Entropy \textbf{15}, 2100 (2013).

\bibitem{chen2016}B. Lin and J. Chen, Performance analysis of an
irreversible quantum heat engine working with harmonic oscillators,
Phys. Rev. E \textbf{67}, 046105 (2003).

\bibitem{chen2002}J. He, J. Chen, and B. Hua, Quantum refrigeration
cycles using spin-1/2 systems as the working substance, Phys. Rev.
E \textbf{65}, 036145 (2002).

\bibitem{Batalh=0000E3o2014}T. B. Batalhão, A. M. Souza, L. Mazzola,
R. Auccaise, R. S. Sarthour, I. S. Oliveira, J. Goold, G. De Chiara,
M. Paternostro, and R. M. Serra, Experimental reconstruction of work
distribution and study of fluctuation relations in a closed quantum
system, Phys. Rev. Lett. \textbf{113}, 140601 (2014).

\bibitem{=0000C5berg2018}J. Åberg, Fully quantum fluctuation theorems,
Phys. Rev. X \textbf{8}, 011019 (2018).

\bibitem{Micadei2020}K. Micadei, G. T. Landi, and E. Lutz, Quantum
fluctuation theorems beyond two-point measurements,  Phys. Rev. Lett.
\textbf{124}, 090602 (2020).

\bibitem{Timpanaro2019}A. M. Timpanaro, G. Guarnieri, J.Goold, and
G. T. Landi, Thermodynamic uncertainty relations from exchange fluctuation
theorems, Phys. Rev. Lett. \textbf{123}, 090604 (2019).

\bibitem{Lee2021}S. Lee, M. Ha, and H. Jeong, Quantumness and thermodynamic
uncertainty relation of the finite-time Otto cycle, Phys. Rev. E \textbf{103},
022136 (2021).

\bibitem{Sacchi2021}M. F. Sacchi, Thermodynamic uncertainty relations
for bosonic Otto engines, Phys. Rev. E \textbf{103}, 012111 (2021).

\bibitem{Zhang2020}Y. Zhang, Optimization performance of quantum
Otto heat engines and refrigerators with squeezed thermal reservoirs,
Physica A \textbf{559}, 125083 (2020).

\bibitem{Huang2012}X. L. Huang, T. Wang, and X. X. Yi, Effects of
reservoir squeezing on quantum systems and work extraction, Phys.
Rev. E \textbf{86}, 051105 (2012).

\bibitem{Scully2003}M. O. Scully, M. S. Zubairy, G. S. Agarwal, and
H. Walther, Extracting work from a single heat bath via vanishing
quantum coherence, Science \textbf{299}, 5608 (2003).

\bibitem{Rodrigues2019}F. L. S. Rodrigues, G. De Chiara, M. Paternostro,
and G. T. Landi, Thermodynamics of weakly coherent collisional models,
Phys. Rev. Lett. \textbf{123}, 140601 (2019).

\bibitem{Cresser2021}J. D. Cresser and J. Anders, Weak and ultrastrong
coupling limits of the quantum mean force Gibbs state, arXiv: 2104.
12606.

\bibitem{Strasberg2016}P. Strasberg, G. Schaller, N. Lambert, and
T. Brandes, Nonequilibrium thermodynamics in the strong coupling and
non-Markovian regime based on a reaction coordinate mapping, New J.
Phys. \textbf{18}, 073007 (2016).

\bibitem{Breuer2009}H. P. Breuer, E.-M. Laine, and J. Piilo, Measure
for the degree of non-Markovian behavior of quantum processes in open
systems, Phys. Rev. Lett. \textbf{103}, 210401 (2009).

\bibitem{Laine2010}E.-M. Laine, J. Piilo, and H.-P. Breuer, Measure
for the non-Markovianity of quantum processes, Phys. Rev. A \textbf{81},
062115 (2010).

\bibitem{Camati2019}P. A. Camati, J. F. G. Santos, and R. M. Serra,
Coherence effects in the performance of the quantum Otto heat engine,
Phys. Rev. A \textbf{99}, 062103 (2019).

\bibitem{Santos2019}J. P. Santos, L. C. Céleri, G. T. Landi, and
M. Paternostro, The role of quantum coherence in non-equilibrium entropy
production, npj Quantum Inf. \textbf{5}, 23 (2019).

\bibitem{Francica2019}G. Francica, J. Goold, and F. Plastina, The
role of coherence in the non-equilibrium thermodynamics of quantum
systems, Phys. Rev. E \textbf{99}, 042105 (2019).

\bibitem{Abiuso2019}P. Abiuso and V. Giovannetti, Non-Markov enhancement
of maximum power for quantum thermal machines, Phys. Rev. A \textbf{99},
052106 (2019).

\bibitem{Shirai2021}Y. Shirai, K. Hashimoto, R. Tezuka, C. Uchiyama,
and N. Hatano, Non-Markovian effect on quantum Otto engine: Role of
system-reservoir interaction, Phys. Rev. Research \textbf{3}, 023078
(2021).

\bibitem{Guarnieri2016}G. Guarnieri, C. Uchiyama, and B. Vacchini,
Energy backflow and non-Markovian dynamics, Phys. Rev. A \textbf{93},
012118 (2016).

\bibitem{Camati2020}P. A. Camati, J. F. G. Santos, and R. M. Serra,
Employing non-Markovian effects to improve the performance of a quantum
Otto refrigerator, Phys. Rev. A \textbf{102}, 012217 (2020).

\bibitem{Pozas-Kerstjens2018}A. Pozas-Kerstjens, E. G Brown, and
K. V. Hovhannisyan, A quantum Otto engine with finite heat baths:
energy, correlations, and degradation, New J. Phys. \textbf{20} 043034
(2018).

\bibitem{Assis2020}R. J. de Assis, J. S. Sales, J. A. R. da Cunha,
and N. G. de Almeida, Universal two-level quantum Otto machine under
a squeezed reservoir, Phys. Rev. E \textbf{102}, 052131 (2020).

\bibitem{Klaers2017}J. Klaers, S. Faelt, A. Imamoglu, and E. Togan,
Squeezed thermal reservoirs as a resource for a nanomechanical engine
beyond the Carnot limit, Phys. Rev. X \textbf{7}, 031044 (2017).

\bibitem{Ro=0000DFnagel2014}J. Roßnagel, O. Abah, F. Schmidt-Kaler,
K. Singer, and E. Lutz, Nanoscale heat engine beyond the Carnot limit,
Phys. Rev. Lett. \textbf{112}, 030602 (2014).

\bibitem{Peterson2019}J. P.\LyXThinSpace S. Peterson, T. B. Batalhão,
M. Herrera, A. M. Souza, R. S. Sarthour, I. S. Oliveira, and R. M.
Serra, Experimental characterization of a spin quantum heat engine,
Phys. Rev. Lett. \textbf{123}, 240601 (2019).

\bibitem{Klatzow2019}J. Klatzow, J. N. Becker, P. M. Ledingham, C.
Weinzetl, K. T. Kaczmarek, D. J. Saunders, J. Nunn, I. A. Walmsley,
R.Uzdin, and E. Poem, Experimental demonstration of quantum effects
in the operation of microscopic heat engines, Phys. Rev. Lett. \textbf{122},
110601 (2019).

\bibitem{Denzler2021}T. Denzler, J. F. G. Santos, E. Lutz, and R.
M. Serra, Nonequilibrium fluctuations of a quantum heat engine, arXiv:
2104. 13427.

\bibitem{Szilard1929}L. Szilard, über die entropieverminderung in
einem thermodynamischen system bei eingriffen intelligenter wesen,
Zeitschrift fur Physik \textbf{53}, 840 (1929).

\bibitem{Leff1990}H.S. Leff and A.F. Rex, \textit{Maxwell\textquoteright s
Demon: Entropy, Information, Computation, Computing} (Princeton University
Press, Princeton, 1990).

\bibitem{Elouard2017}C. Elouard, D. Herrera-Martí, B. Huard, and
A. Auffèves, Extracting work from quantum measurement in Maxwell\textquoteright s
Demon engines, Phys. Rev. Lett. \textbf{118}, 260603 (2017).

\bibitem{Park2013}J. J. Park, K.-H. Kim, T. Sagawa, and S. W.Kim,
Heat engine driven by purely quantum information, Phys. Rev. Lett.
\textbf{111}, 230402 (2013).

\bibitem{Toyabe2010}S. Toyabe, T. Sagawa, M. Ueda, E. Muneyuki, and
M. Sano, Information heat engine: converting information to energy
by feedback control, Nat. Phys. \textbf{6}, 988-992 (2010).

\bibitem{Yi2017}J. Yi, P. Talkner, and Y. W. Kim, Single-temperature
quantum engine without feedback control, Phys. Rev. E \textbf{96},
022108 (2017).

\bibitem{Ding2018}X. Ding, J. Yi, Y. W. Kim, and P. Talkner, Measurement
driven single temperature engine, Phys. Rev. E \textbf{98}, 042122
(2018).

\bibitem{Behzadi2020}N. Behzadi, Quantum engine based on general
measurements, J. Phys. A: Math. Theor. \textbf{54} 015304 (2020).

\bibitem{Elouard2018}C. Elouard and A. N. Jordan, Efficient quantum
measurement engines, Phys. Rev. Lett. \textbf{120}, 260601 (2018).

\bibitem{Bresque2021}L. Bresque, P. A. Camati, S. Rogers, K. Murch,
A. N. Jordan, and A. Auffèves, Two-qubit engine fueled by entanglement
and local measurements, Phys. Rev. Lett. \textbf{126}, 120605 (2021).

\bibitem{Brandner2015}K. Brandner, M. Bauer, M.\LyXThinSpace T. Schmid,
and U. Seifert, Coherence-enhanced efficiency of feedback-driven quantum
engines, New J. Phys. \textbf{17},\textbf{ }065006 (2015).

\bibitem{Aharonov1988}Y. Aharonov, D. Z. Albert, and L. Vaidman,
How the result of a measurement of a component of the spin of a spin-1/2
particle can turn out to be 100, Phys. Rev. Lett. \textbf{60} 1351-1354
(1988).

\bibitem{Duck1989}I. M. Duck, P. M. Stevenson, and E. C. G. Sudarshan,
The sense in which a \textquotedbl weak measurement\textquotedbl{}
of a spin-\textonehalf{} particle's spin component yields a value
100, Phys. Rev. D \textbf{40}, 2112 (1989).

\bibitem{Flack2014}R. Flack and B. J. Hiley, Weak measurement and
its experimental realisation, J. Phys.: Conf. Ser. \textbf{504}, 012016
(2014).

\bibitem{Hosten2008}O. Hosten and P. Kwiat, Observation of the spin
hall effect of light via weak measurements, Science \textbf{319},
787 (2008).

\bibitem{Lundeen2011}J. S. Lundeen, B. Sutherland, A. Patel, C. Stewart,
and C. Bamber, Direct measurement of the quantum wavefunction, Nature
\textbf{474}, 188 (2011).

\bibitem{Monroe2021}J. T. Monroe, N. Yunger Halpern, T.Lee, and K.
W. Murch, Weak measurement of a superconducting qubit reconciles incompatible
operators, Phys. Rev. Lett. \textbf{126}, 100403 (2021).

\bibitem{Jacobs2014}K. Jacobs, \textit{Quantum measurements theory
and its applications} (Cambridge University Press, Cambridge, England,
2014).

\bibitem{Wiseman2010}H. M. Wiseman and G. J. Milburn, \textit{Quantum
measurement and control} (Cambridge University Press, Cambridge, England,
2010).

\bibitem{Zagoskin2012}A. M. Zagoskin, S. Savel\textquoteright ev,
Franco Nori, and F. V. Kusmartsev, Squeezing as the source of inefficiency
in the quantum Otto cycle, Phys. Rev. B \textbf{86}, 014501 (2012).

\bibitem{Rezek2010}Y. Rezek, Reflections on friction in quantum mechanics,
Entropy \textbf{12}, 1885 (2010).

\bibitem{Feldmann2006}T. Feldmann and R. Kosloff, Quantum lubrication:
Suppression of friction in a first-principles four-stroke heat engine,
Phys. Rev. E \textbf{73}, 025107(R) (2006).

\bibitem{Kosloff2002}R. Kosloff and T. Feldmann, Discrete four-stroke
quantum heat engine exploring the origin of friction, Phys. Rev. E
\textbf{65}, 055102(R) (2002).

\bibitem{Oliveira_Book2007}I. S. Oliveira, T. J. Bonagamba, R. S.
Sarthour, J. C. C. Freitas, and E. R. deAzevedo, \textit{NMR quantum
information processing} (Elsevier, Amsterdam, 2007).

\bibitem{Quan2007}T. Quan, Y. Liu, C. P. Sun, and F. Nori, Quantum
thermodynamic cycles and quantum heat engines, Phys. Rev. E \textbf{76},
031105 (2007).

\bibitem{Chen2018}J. Chen, H. Dong, and C. Sun, Bose-Fermi duality
in a quantum Otto heat engine with trapped repulsive bosons, Phys.
Rev. E \textbf{98}, 062119 (2018).

\bibitem{Manzano2018}G. Manzano, F. Plastina, and R. Zambrini, Optimal
work extraction and thermodynamics of quantum measurements and correlations,
Phys. Rev. Lett. \textbf{121}, 120602 (2018).

\bibitem{Lloyd1997}S. Lloyd, Quantum-mechanical Maxwell's demon,
Phys. Rev. A \textbf{56}, 3374 (1997).

\bibitem{Campisi2011}M. Campisi, P. Hänggi, and P. Talkner, Colloquium:
Quantum fluctuation relations: Foundations and applications, Rev.
Mod. Phys. \textbf{83}, 771 (2011).

\bibitem{Ehrich2020}J. Ehrich, M. Esposito, F. Barra, and J. M. R.
Parrondo, Micro-reversibility and thermalization with collisional
baths, Physica A \textbf{552}, 122108 (2020).
\end{thebibliography}
\end{document}